# Phase field theory of polycrystalline solidification in three dimensions

T. PUSZTAI, G. BORTEL and L. GRÁNÁSY
*Research Institute for Solid State Physics and Optics*
*H-1525 Budapest, POB 49, Hungary*

**Abstract.** – A phase field theory of polycrystalline solidification is presented that is able to describe the nucleation and growth of anisotropic particles with different crystallographic orientation in 3D dimensions. As opposed with the two-dimensional case, where a single orientation field suffices, in three dimensions, minimum three fields are needed. The free energy of grain boundaries is assumed to be proportional to the angular difference between the adjacent crystals expressed here in terms of the differences of the four symmetric Euler parameters. The equations of motion for these fields are obtained from variational principles. Illustrative calculations are performed for polycrystalline solidification with dendritic, needle and spherulitic growth morphologies.

*Introduction.* – One of the greatest challenges to computational materials science is the modeling of polycrystalline solidification. The main source of difficulties is that in a general case, the nucleation of crystallites of different crystallographic orientations needs to be incorporated into the model. Recently, we developed a phase field theory that has successfully addressed this problem in two-dimensional (2D) and quasi-two dimensional systems [1]. Our approach describes the formation of such complex polycrystalline growth patterns as the disordered ("dizzy") dendrites [2,3], fractallike aggregates [2], spherulites [2-4] and 'quadrites' [2,4] observed in polymeric thin films. (Similar models have been used to describe grain boundary dynamics, wetting and grain rotation [5-7].) The polycrystalline structures are of vast importance for technical alloys, polymers, minerals, and have also biological relevance (e.g., semicrystalline amyloid spherulites have been associated with the Alzheimer and Kreutzfeld-Jacob diseases, and the type II. diabetes [8]). Simulation of complex polycrystalline growth patterns is restricted to two dimensions as yet. Generalization of these models to three dimensions is not trivial. While in two dimensions a single field is sufficient to describe the local crystallographic orientation, relative to which the anisotropies of the interfacial and dynamic properties are counted, in three dimensions at least three parameter fields (e.g., the Euler angles) are needed to formulate the theory. Herein, we present a phase field theory for polycrystalline freezing in three dimensions, and a few illustrative simulations.

*Theory.* – As in 2D, the free energy functional we propose consists of the usual square-gradient, double well, and driving force terms, and an orientational contribution $f_{ori}$:

$$F = \int d^3r \left\{ \frac{\varepsilon_\phi^2 T}{2} |\nabla \phi|^2 + f(\phi, c, T) + f_{ori} \right\}, \quad (1)$$

where the local phase state of matter (solid or liquid) is characterized by the phase field $\phi$ (a structural order parameter) and the solute concentration $c$, $\varepsilon_\phi$ is a constant, $T$ is the temperature. The gradient term for the phase field leads to a diffuse crystal-liquid interface, a feature observed both in experiment [9] and computer simulations [10]. The local free energy density is assumed to have the form $f(\phi, c, T) = w(\phi) T g(\phi) + [1 - p(\phi)] f_S(c) + p(\phi) f_L(c)$, where the "double well" and "interpolation" functions have the forms $g(\phi) = \frac{1}{4} \phi^2 (1 - \phi)^2$ and $p(\phi) = \phi^3 (10 - 15\phi + 6\phi^2)$, respectively, while the free energy scale $w(\phi) = (1 - c) w_A + c w_B$ [11,12]. The respective free energy surface has two minima ($\phi = 0$ and $\phi = 1$, corresponding to the crystalline and liquid phases), whose relative depth is the driving force for crystallization and is a function of both temperature and composition as specified by the free energy densities in the bulk solid and liquid, $f_{S,L}(c,T)$, taken here from the ideal solution model. (For recent review on details of the phase field technique see [2,13,14].)

The orientational contribution $f_{ori}$ should be invariant to rotations of the whole system, and has to penalize spatial changes in the crystal orientation. We define this term analogously to the 2D case [2,3]. First, we consider two semi-infinite crystals of identical structure in contact (a bi-crystal). According to experiment, the grain boundary that separates them is localized on the nanometer scale, and the free energy associated with small-angle grain boundaries increases approximately linearly with the misorientation of the two crystals, saturating at about twice of the free energy of the solid-liquid interface. Our goal is to reproduce reasonably, the behavior of small angle grain boundaries. First, we define the norm of the misorientation of the two crystals.

The *local crystallographic* orientation is considered as the relative orientation of a local coordinate system (fixed to the crystal lattice) with respect to a reference system (fixed to the laboratory frame). This relative orientation is uniquely defined by a single rotation of angle $\delta$ around a specific axis, and can be expressed in terms of the three Euler angles. However, this representation has disadvantages: It has divergences at the poles $\theta = 0$ and $\pi$, and one has to use trigonometric functions that are time consuming in numerical calculations. Therefore, we opt for the four symmetric Euler parameters, $q_0 = \cos(\delta/2)$, $q_1 = c_1 \sin(\delta/2)$, $q_2 = c_2 \sin(\delta/2)$, and $q_3 = c_3 \sin(\delta/2)$, a representation free of such difficulties. (Here $c_i$ are the components of the unit vector $c$ of the rotation axis.) These four parameters $q = (q_0, q_1, q_2, q_3)$, are sometimes referred to as *quaternions*, they satisfy the relationship $\sum_i q_i^2 = 1$, therefore, can be viewed as a point of the four-dimensional (4D) unit sphere [15]. (Here $\sum_i$ stands for summation with respect to $i = 0, 1, 2,$ and 3, a notation used throughout this paper.)

The angular difference $\delta$ between two orientations represented by quaternions $q_1$ and $q_2$ can be expressed as $\cos(\delta) = \frac{1}{2} [\text{Tr}(\mathbf{R}) - 1]$, where the matrix of rotation $\mathbf{R}$ is related to the individual rotation matrices $\mathbf{R}(q_1)$ and $\mathbf{R}(q_2)$, that rotate the reference system into the corresponding local orientations, as $\mathbf{R} = \mathbf{R}(q_1) \cdot \mathbf{R}(q_2)^{-1}$ (see fig. 1). After lengthy but straightforward algebraic manipulations one finds that the angular difference can be expressed in terms of the differences of quaternion coordinates: $\cos(\delta) = 1 - 2\Delta^2 + \Delta^4/2$, where $\Delta^2 = (q_2 - q_1)^2 = \sum_i \Delta q_i^2$, is the square of the Euclidian distance between the points $q_1$ and $q_2$ on the 4D unit sphere. Comparing this expression with the Taylor expansion of the function $\cos(\delta)$, one finds that $2\Delta$ is indeed an excellent approximation of $\delta$. Relying on this approximation, we express the orientational difference as $\delta \approx 2\Delta$.

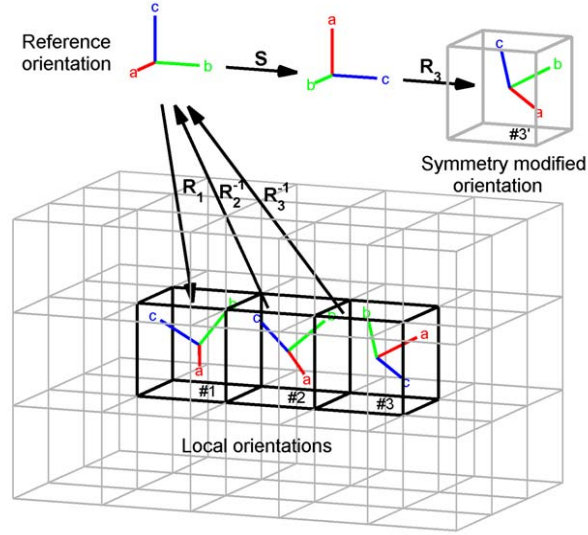

Fig. 1 – Calculation of misorientation with and without crystal symmetries. For the cells #1 and #2 $\delta$ is calculated from the trace of $\mathbf{R}(\mathbf{q}_1)\cdot\mathbf{R}(\mathbf{q}_2)^{-1}$. The cells on the right (#2 and #3) illustrate how crystal symmetries can be taken into account. First the local coordinate system of cell #3 is transformed temporarily into a physically equivalent orientation, #3' with the help of an appropriately selected symmetry operation, $\mathbf{S}$ (see text). Then the angular difference $\delta$ between cells #3 and #2 is determined by $\mathbf{R}(\mathbf{q}_3)\cdot\mathbf{S}\cdot\mathbf{R}(\mathbf{q}_2)^{-1}$.

To model the behavior expected for low angle grain boundaries, we assume that the free energy of a bi-crystal due to the grain boundary is proportional to the misorientation angle: $\gamma_{gb} \propto |\delta| \approx 2\{\sum_i \Delta q_i^2\}^{1/2}$. In accordance, we postulate that the density of the orientational free energy reads as

$$f_{ori} = 2HT[1 - p(\phi)]\left\{\sum_i \left(\nabla q_i^2\right)\right\}^{1/2}, \qquad (2)$$

where the grain boundary energy scales with the constant $H$. This definition boils down to the expression used in the two-dimensional model $f_{ori} = HT[1 - p(\phi)]|\nabla\delta|$ [1-4], provided that the orientational transition across the grain boundary has a fixed rotation axis as in 2D.

To model crystal nucleation in the liquid, this "vectorial" orientation field $\mathbf{q}(\mathbf{r})$ is extended to the liquid, where it is made to fluctuate in time and space as in 2D [1-4]. Assigning local crystal orientation to liquid regions, even a fluctuating one, may seem artificial at first sight. However, due to geometrical and/or chemical constraints, a short-range order exists even in simple liquids, which is often similar to the one in the solid. Rotating the crystalline first-neighbor shell so that it aligns optimally with the local liquid structure, one may assign a local orientation to every atom in the liquid. The orientation obtained in this manner fluctuates in time and space. The correlation of the atomic positions/angles shows how good this fit is. (In the model, the fluctuating orientation fields and the phase field play these roles.) Approaching the solid from the liquid, the orientation becomes more definite (the amplitude of the orientational fluctuations decreases) and matches to that of the solid, while the correlation between the local liquid structure and the crystal structure improves. The present $f_{ori}$ recovers this behavior by realizing a strong coupling between the orientation and phase fields.

$f_{ori}$ is consists of the factor $(1 - p(\phi))$ to avoid double counting the orientational contribution in the liquid, which is *per definitionem* incorporated into the free energy of the bulk liquid. With appropriate choice of the model parameters, an ordered liquid layer surrounds the crystal as seen in atomistic simulations [10].

Time evolution is governed by relaxational dynamics and noise terms are added to model thermal fluctuations. The equations of motion for the phase and concentration fields are straightforward [1-4,13,14,16,17]. In contrast, we take into account the quaternion properties ($\sum_i q_i^2 = 1$) in the derivation of the equation of motion for the four orientational fields $q_i(\mathbf{r})$ using the method of Lagrange multipliers,

$$\frac{\partial q_i}{\partial t} = -M_q \frac{\delta F}{\delta q_i} + \zeta_i = M_q \left\{ \nabla \left( \frac{\partial I}{\partial \nabla q_i} \right) - \frac{\partial I}{\partial q_i} \right\} + \zeta_i \qquad (4)$$

where $M_q$ is the common mobility coefficient for the symmetric quaternion fields, $I$ is the total free energy density (including the gradient terms and the term with the Lagrange multiplier), while $\zeta_i$ are the appropriate noise terms. Utilizing the relationship $\sum_i q_i (\partial q_i/\partial t) = 0$ that follows from the quaternion properties, and expressing the Lagrange multiplier in terms of $q_i$ and $\nabla q_i$, the equation of motion for the orientation (quaternion) fields can be expressed as

$$\frac{\partial q_i}{\partial t} = M_q \left\{ \begin{array}{l} \nabla \left( HT[1-p(\phi)] \frac{\nabla q_i}{\left[\sum_l (\nabla q_l)^2\right]^{1/2}} \right) \\ -q_i \sum_k q_k \nabla \left( HT[1-p(\phi)] \frac{\nabla q_k}{\left[\sum_l (\nabla q_l)^2\right]^{1/2}} \right) \end{array} \right\} + \zeta_i. \qquad (5)$$

Gaussian white noises of amplitude $\zeta_i = \zeta_{S,i} + (\zeta_{L,i} - \zeta_{S,i}) p(\phi)$ have been added to the orientation fields so that the quaternion properties of the $q_i$ fields are retained. ($\zeta_{L,i}$ and $\zeta_{S,i}$ are the amplitudes in the liquid and solid.)

This formulation of the model is valid for triclinic lattice without symmetries (space group P1). In the case of other crystals, the crystal symmetries yield equivalent orientations that do not form grain boundaries. These symmetries can be taken into account, when discretizing the differential operators used in the equations of motions for the quaternion fields. Calculating the angular difference between a central cell and its neighbors, all equivalent orientations of the neighbor have to be considered, the respective angular differences $\delta$ be calculated (using matrices of rotation $\mathbf{R} = \mathbf{R}(\mathbf{q_3}) \cdot \mathbf{S} \cdot \mathbf{R}(\mathbf{q_2})^{-1}$, where $\mathbf{S}$ is a symmetry operator, see fig. 1), of which the smallest $\delta$ value shall be used in calculating the differential operator. All the present simulations have been performed assuming a triclinic structure.

The equations of motion have been solved numerically using an explicit finite difference scheme. Periodic boundary conditions were applied. The time and spatial steps were chosen to ensure stability of our solutions. The noise has been discretized as described by Karma and Rappel [18] and Plapp [19]. As the size of the three-dimensional simulation box and the accessible time window are seriously restricted by the computational power needed, to observe nucleation, we enhanced the amplitude of the noise relative to the value from the fluctuation-dissipation theorem. A parallel code has been developed and run on two PC clusters, consisting of 60 nodes and a server machine each.

In the present computations, we use the physical properties of the Ni-Cu system [11]. This choice is not particularly restrictive, as it is formally equivalent to a pure material, where thermal diffusion replaces solute diffusion as the dominant transport mechanism [11]. Note that our model is no way restricted to metals. We fix the temperature to be 1574 K, as in previous studies. Dendritic growth in our simulations originates from the anisotropy of the phase field mobility $M_\phi = M_{\phi,0}\{1 - 3\gamma + 4\gamma$

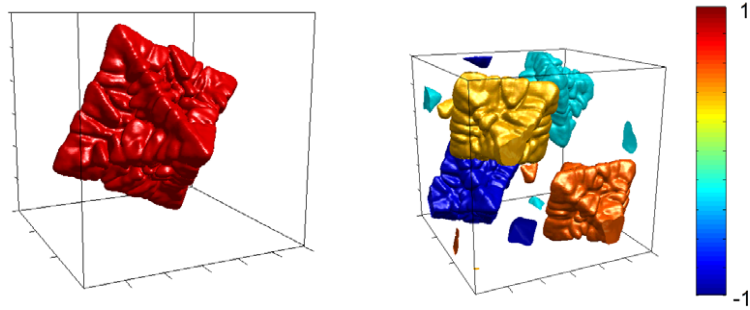

Fig. 2 – Growth of a single (left) and four (right) three-dimensional dendritic particles with random orientation on a 300×300×300 grid. (Note the effect of periodic boundary condition on the pattern in the latter case. The branches growing out of the simulation window on one side enter on the opposite side.) The color bar shows the value of the 0$^{th}$ quaternion field on the surfaces corresponding to $\phi = 0.5$

$[(\nabla\phi)_x^4 + (\nabla\phi)_y^4 + (\nabla\phi)_z^4]/|\nabla\phi|^4\}$ that reflects the orientation dependence of the molecular attachment kinetics [14]. In the simulations for needle crystals, a rotational ellipsoidal symmetry of the phase field mobility has been assumed. Our calculations were performed at supersaturations $S = (c_L - c)/(c_L - c_S) = 0.78$ for dendritic and needle crystals, and at $S = 1.0$ for the polycrystalline spherulites. Here $c_L = 0.466219$, $c_S = 0.399112$ and c are the concentrations at the liquidus, solidus, and the initial homogeneous liquid mixture, respectively.

Since the physical thickness of the interface is in the nanometer range and the typical solidification structures are far larger (μm to mm), a full simulation of polycrystalline solidification from nucleation to particle impingement cannot be performed even with the fastest of the present supercomputers. Since we seek here a qualitative understanding, following previous work [1-4,12,20], the interface thickness has been increased (by a factor of about 20, $d = 20.6$ nm), while the interface free energy of the pure constituents has been reduced (by a factor of 6). This allows us to follow the life of crystallites from birth to impingement on each other. The time and spatial steps were $\Delta t = 1.31$ ns and $\Delta x = 13.1$ nm. The diffusion coefficient in the liquid was $D_L = 10^{-9}$ m$^2$/s. Unless stated otherwise, dimensionless mobilities of $M_{\phi,0L} = 3.55\times 10^{-1}$ m$^3$/Js and $M_{q,L} = 8.17$ m$^3$/Js, and $M_{q,S} = 0$ were applied, while $D_S = 0$ was taken in the solid. Gaussian white noises of amplitude from the fluctuation-dissipation theorem were used, except in the nucleation runs, where the phase field noise was enhanced by a factor of 2.3 to speed up the process.

*Results.* – We performed illustrative simulations for anisotropic growth of crystals. Dendritic solidification of a single particle and four particles with random orientation are shown in fig. 2. These morphologies are consistent with previous results for single crystal growth [14,16-18].

Snapshots showing the time evolution of noise-induced nucleation and chemical-diffusion-controlled growth of needle crystals, often termed "soft-impingement", are presented in fig. 3 together with the time dependence of the crystalline fraction $X(t)$, the Avrami plot $\ln\{-\ln[1 - X(t)/X_{max}]\}$ vs. $\ln(t)$, and the Avrami-Kolmogorov exponent $p$ (slope of the Avrami plot) that characterizes the mechanism of transformation as a function of crystalline fraction. It is found that, in agreement with previous simulation results in 2D [1,2,21] and experiment on crystallization of metallic glasses [22], $p$ decreases with the transformed fraction.

Reducing the orientational mobility, a uniform crystallographic orientation cannot be established at the growth front, and new crystal grains form, that are induced by orientational defects frozen into the solid [2,4]. At large driving forces space filling polycrystalline forms called spherulites and seaweed appear [2,23]. A three-dimensional polycrystalline spherulite, a polycrystalline seaweed and its grain structure are displayed in fig. 4.

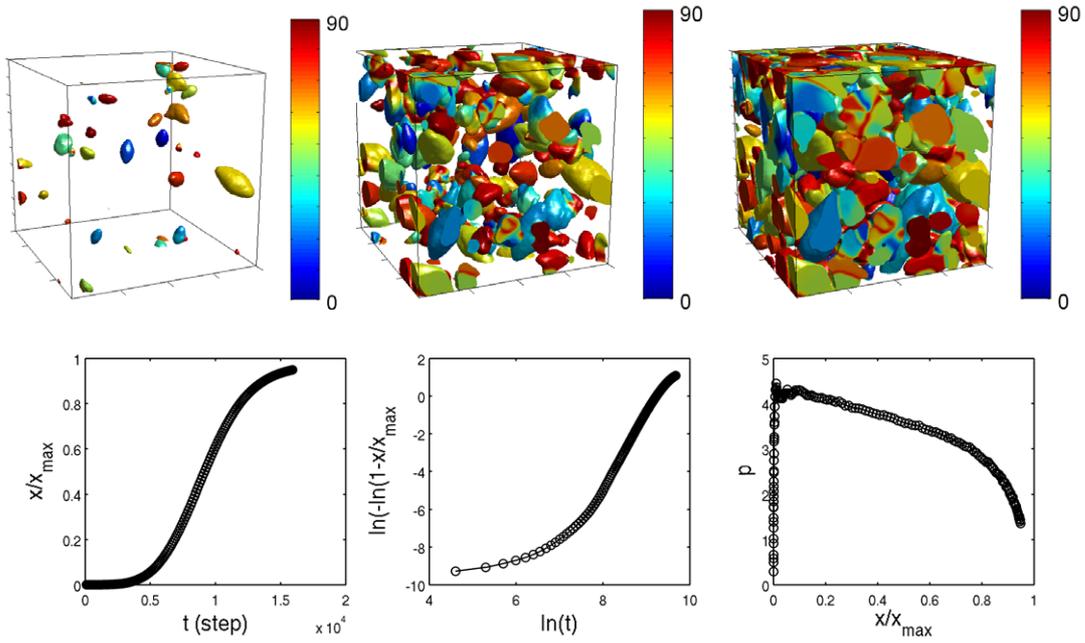

Fig. 3 – Nucleation and growth of needle crystals in the phase field theory on a 200×200×200 grid (upper row). The growth fronts are colored according to the angular difference between the *z* axes of the local and laboratory frames. Panels in the lower row show the time dependence of the crystalline fraction, Avrami plot, and the Avrami-Kolmogorov exponent as a function of transformed fraction, respectively.

Summarizing, we have demonstrated that our phase field method is capable for describing various polycrystalline solidification processes including impinging dendritic particles, needle crystals nucleating with random orientation, and spherulitic polycrystals formed by growth front nucleation. These results are consistent with experiment and with recent simulations in two dimensions. Our method opens up the way for phase field modeling of complex polycrystalline morphologies in three dimensions. A more complete study of the three-dimensional polycrystalline structures is under way.

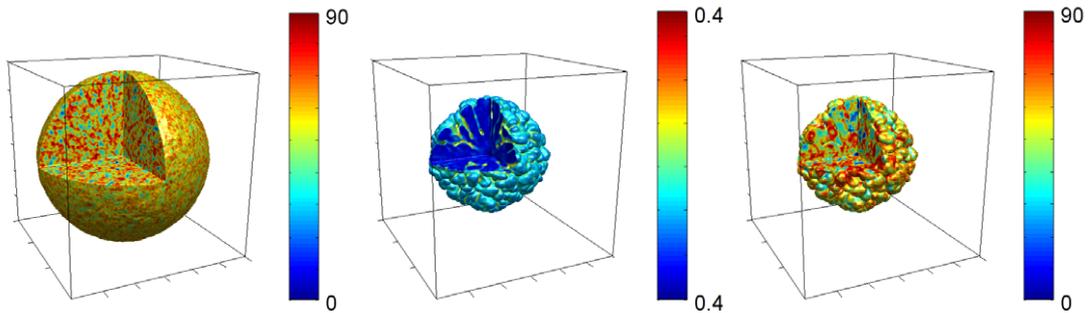

Fig. 4 – Polycrystalline spherulitic (left) and seaweed (center and right) patterns grown in the phase field theory on a 300×300×300 grid. $M_{q,L}$ has been reduced by a factor of 3 and 3.5, respectively. Coloring is the same as for fig. 3, except for the central panel, where the concentration field is shown at the surface of the crystal ($\phi = 0.5$).

***

The authors thank M. Tegze for the enlightening discussions. This work has been supported by contracts OTKA-T-037323, ESA PECS Contract No. 98005, and by the EU FP6 Integrated Project IMPRESS, and forms part of the ESA MAP Projects No. AO-99-101 and AO-99-114. T. P. and G. B. acknowledge support by the Bolyai János Scholarship of the Hungarian Academy of Sciences.